\documentclass[twoside,twocolumn,9pt]{article}
\usepackage{extsizes}
\usepackage[super,sort&compress,comma]{natbib} 
\usepackage[version=3]{mhchem}
\usepackage[left=1.5cm, right=1.5cm, top=1.785cm, bottom=2.0cm]{geometry}
\usepackage{balance}
\usepackage{mathptmx}
\usepackage{sectsty}
\usepackage{graphicx} 
\graphicspath{{../Bilder/}{../Bilder/pdf/}}
\usepackage{lastpage}
\usepackage[format=plain,justification=justified,singlelinecheck=false,font={stretch=1.125,small,sf},labelfont=bf,labelsep=space]{caption}
\usepackage{float}
\usepackage{fancyhdr}
\usepackage{fnpos}
\usepackage[english]{babel}
\addto{\captionsenglish}{%
  
}
\usepackage{array}
\usepackage{droidsans}
\usepackage{charter}
\usepackage[T1]{fontenc}
\usepackage[usenames,dvipsnames]{xcolor}
\usepackage{setspace}
\usepackage[compact]{titlesec}
\usepackage{hyperref}
\usepackage{todonotes}
\usepackage{ulem}

\usepackage{epstopdf}%This line makes .eps figures into .pdf - please comment out if not required.

\definecolor{cream}{RGB}{222,217,201}

\begin{document}

\pagestyle{fancy}
\thispagestyle{plain}
\fancypagestyle{plain}{
%%%HEADER%%%
\renewcommand{\headrulewidth}{0pt}
}
%%%END OF HEADER%%%

%%%PAGE SETUP - Please do not change any commands within this section%%%
\makeFNbottom
\makeatletter
\renewcommand\LARGE{\@setfontsize\LARGE{15pt}{17}}
\renewcommand\Large{\@setfontsize\Large{12pt}{14}}
\renewcommand\large{\@setfontsize\large{10pt}{12}}
\renewcommand\footnotesize{\@setfontsize\footnotesize{7pt}{10}}
\makeatother

\renewcommand{\thefootnote}{\fnsymbol{footnote}}
\renewcommand\footnoterule{\vspace*{1pt}% 
\color{cream}\hrule width 3.5in height 0.4pt \color{black}\vspace*{5pt}} 
\setcounter{secnumdepth}{5}

\makeatletter 
\renewcommand\@biblabel[1]{#1}            
\renewcommand\@makefntext[1]% 
{\noindent\makebox[0pt][r]{\@thefnmark\,}#1}
\makeatother 
\renewcommand{\figurename}{\small{Fig.}~}
\sectionfont{\sffamily\Large}
\subsectionfont{\normalsize}
\subsubsectionfont{\bf}
\setstretch{1.125} %In particular, please do not alter this line.
\setlength{\skip\footins}{0.8cm}
\setlength{\footnotesep}{0.25cm}
\setlength{\jot}{10pt}
\titlespacing*{\section}{0pt}{4pt}{4pt}
\titlespacing*{\subsection}{0pt}{15pt}{1pt}
%%%END OF PAGE SETUP%%%

%%%FOOTER%%%
\fancyfoot{}
\fancyfoot[LO,RE]{\vspace{-7.1pt}}%\includegraphics[height=9pt]{head_foot/LF}}
\fancyfoot[CO]{\vspace{-7.1pt}\hspace{13.2cm}}%\includegraphics{head_foot/RF}}
\fancyfoot[CE]{\vspace{-7.2pt}\hspace{-14.2cm}}%\includegraphics{head_foot/RF}}
\fancyfoot[RO]{\footnotesize{\sffamily{1--\pageref{LastPage} ~\textbar  \hspace{2pt}\thepage}}}
\fancyfoot[LE]{\footnotesize{\sffamily{\thepage~\textbar\hspace{3.45cm} 1--\pageref{LastPage}}}}
\fancyhead{}
\renewcommand{\headrulewidth}{0pt} 
\renewcommand{\footrulewidth}{0pt}
\setlength{\arrayrulewidth}{1pt}
\setlength{\columnsep}{6.5mm}
\setlength\bibsep{1pt}
%%%END OF FOOTER%%%

%%%FIGURE SETUP - please do not change any commands within this section%%%
\makeatletter 
\newlength{\figrulesep} 
\setlength{\figrulesep}{0.5\textfloatsep} 

\newcommand{\topfigrule}{\vspace*{-1pt}% 
\noindent{\color{cream}\rule[-\figrulesep]{\columnwidth}{1.5pt}} }

\newcommand{\botfigrule}{\vspace*{-2pt}% 
\noindent{\color{cream}\rule[\figrulesep]{\columnwidth}{1.5pt}} }

\newcommand{\dblfigrule}{\vspace*{-1pt}% 
\noindent{\color{cream}\rule[-\figrulesep]{\textwidth}{1.5pt}} }

\makeatother
%%%END OF FIGURE SETUP%%%

%%%TITLE, AUTHORS AND ABSTRACT%%%
\twocolumn[
  \begin{@twocolumnfalse}
%{\includegraphics[height=30pt]{head_foot/journal_name}\hfill\raisebox{0pt}[0pt][0pt]{\includegraphics[height=55pt]{head_foot/RSC_LOGO_CMYK}}\\[1ex]
%\includegraphics[width=18.5cm]{head_foot/header_bar}}\par
\vspace{1em}
\sffamily
%\begin{tabular}{m{4.5cm} p{13.5cm} }

  %\includegraphics{head_foot/DOI} &
  %\noindent
  \LARGE{\textbf{Dilute gel networks vs. clumpy gels in colloid-polymer mixtures}} \\%Article title goes here instead of the text "This is the title"
\vspace{0.3cm} %& \vspace{0.3cm} \\

%&
%\noindent
\large{M. Gimperlein\,\textit{$^{a}$}, Jasper N. Immink\textit{$^{bc}$} and M. Schmiedeberg\,\textit{$^{d}$}} \\%Author names go here instead of "Full name", etc.

\noindent\normalsize{ Using Brownian dynamics simulations we study gel-forming colloid-polymer mixtures. The focus of this article lies on the differences of dense and dilute gel networks in terms of structure formation both on a local and a global level. We apply reduction algorithms and observe that dilute networks and dense gels differ in the way structural properties like the thickness of strands emerge. We also analyze the percolation behavior and find that two different regimes of percolation exist which might be responsible for structural differences.
  In dilute networks we confirm that solidity is mainly a consequence of pentagonal bipyramids forming in the network. In dense gels also tetrahedral structures influence solidity.} \\%The abstrast goes here instead of the text "The abstract should be..."

%\end{tabular}

\end{@twocolumnfalse}
\vspace{0.6cm}

  ]
%%%END OF TITLE, AUTHORS AND ABSTRACT%%%

%%%FONT SETUP - please do not change any commands within this section
\renewcommand*\rmdefault{bch}\normalfont\upshape
\rmfamily
\section*{}
\vspace{-1cm}

%%%FOOTNOTES%%%

\footnotetext{\textit{$^{a}$~Institut f\"ur Theoretische Physik 1, Friedrich-Alexander-Universit\"at Erlangen-N\"urnberg, D-91058 Erlangen, Germany, E-mail: matthias.gimperlein@fau.de}}
\footnotetext{\textit{$^{b}$~ Condensed Matter Physics Laboratory, Heinrich-Heine-Universit\"at Düsseldorf, D-40225 Düsseldorf, Germany }}
\footnotetext{\textit{$^{c}$~ KWR Water Research Institute, NL-3433 PE Nieuwegein, the Netherlands}}
\footnotetext{\textit{$^{d}$~Institut f\"ur Theoretische Physik 1, Friedrich-Alexander-Universit\"at Erlangen-N\"urnberg, D-91058 Erlangen, Germany, E-mail: michael.schmiedeberg@fau.de}}

%Please use \dag to cite the ESI in the main text of the article.
%If you article does not have ESI please remove the the \dag symbol from the title and the footnotetext below.
%\footnotetext{\dag~Electronic Supplementary Information (ESI) available: [details of any supplementary information available should be included here]. See DOI: 00.0000/00000000.}
%additional addresses can be cited as above using the lower-case letters, c, d, e... If all authors are from the same address, no letter is required

%%%END OF FOOTNOTES%%%

%%%MAIN TEXT%%%%
\section{Introduction}
\label{sec1}
Gelation is an intensively studied phenomenon. A simple system to explore gelation is a colloid-polymer mixture, where there is a phase separation of a dilute and a dense phase. Gelation is observed in the coexistence phase. The depletion attractions between the colloids, that are mediated by the polymers, are usually modelled by AO-interactions\cite{ao,vrij,binder} along with short ranged repulsions and sometimes longer-ranged screened Coulomb repulsions\cite{dlvo1,dlvo2} lead to the formation of numerous complex, partially or completely ordered structures and to a drastic slowdown of dynamics in the system.\cite{verhaegh, lu, speck1, verhaegh2, tanaka3, zaccarelli, kohl}

To be specific, the competition of the attraction and the short-ranged repulsion causes the phase separated regime and the additional longer-ranged repulsions leads to the formation of even more complex heterogeneous structures in this part of the phase diagram.\cite{helgeson, tanaka3}
In experiments the competition of the different interactions can be controlled, e.g., by varying the polymer size and concentration to change the attractive forces or by increasing the salt concentration in the system such that the repulsive Coulomb force is decreased due to additional screening.\cite{kohl}

In the rich phase behavior the following structures are observed: Homogeneous fluids for attractions that are weaker than at the phase separating binodal line and a large zoo of heterogeneous structures for stronger attractions. Close to the binodal line or in case of purely attractive interactions cluster fluids are found\cite{lu2}, which consist of connected particle clusters surrounded by single, non-connected colloidal particles. For systems deep inside the phase separated region, gel networks that consist of thin, long strands of particles are found.\cite{archer, toledano, zhang, mani, helgeson} Depending on the attractive strength, these can be either weakly or strongly connected as shown in our recent article.\cite{gimperlein}

Our main interest in this article are the structural differences between dilute gel networks at low densities and clumpy gels that occur at larger packing  fractions.\\
In dilute systems, except for very high interparticle attraction strengths, particles can almost move freely during most of the equilibration process, i.e., there are no stresses due to percolating strands.  Percolation might only set in at later times of the evolution. In contrast the particles in dense systems percolate nearly instantaneously. Percolation constraints then lead to mechanical forces that modify the structural formation. \cite{tanaka2, helgeson} It was also observed that directed pecolation slows down the relaxation dramatically \cite{kohl}for a system deep in the phase seprated phase\cite{schmiedeberg3} and that it changes the behavior of the gel if subjected to shear.\cite{schmiedeberg4} Here, we investigate how percolation constraints influence the global as well as the local structure formation. In recent research on local structure formation in dilute gel networks it was found that hierarchical formation of local structures is responsible for macroscopic properties of gels.\cite{tanaka} In addition to our study of the differences of dilute and dense gel structures, we use our computer simulations to extend the recent experimental findings on the hierarchical gel formation\cite{tanaka} to dense gels with both weakly and strongly bound particles.

Our article is structured as follows. In section \ref{sec2} we introduce the investigated model system, the simulation procedure, and the methods used to calculate gel network skeletons and distribution functions. In section \ref{sec3} we show our results on structural differences for dilute and dense networks. First we concentrate on skeletonized networks and present results on global properties like thickness of strings, tortuosity and percolation behavior. We find that dilute and dense networks differ in their global properties and that their may be a smooth transition in the phase diagram at around $\varphi=0.1-0.15$. The second part of the results section is dedicated to local structure formation extending the research of Tanaka et al. \cite{tanaka} to higher packing fractions and different interaction strengths. Here we also show that packing fraction and interaction strength have impact on the structure formation process. Before we conclude in section \ref{sec4} we investigate the impact of packing fraction and attraction strength on the onset of solidity by comparing local structures and the average coordination number of the system.

\section{Methods and Simulation procedure}
\label{sec2}

\begin{figure*}
\includegraphics[trim=0cm 0cm 3.5cm 0cm, clip, width=\textwidth]{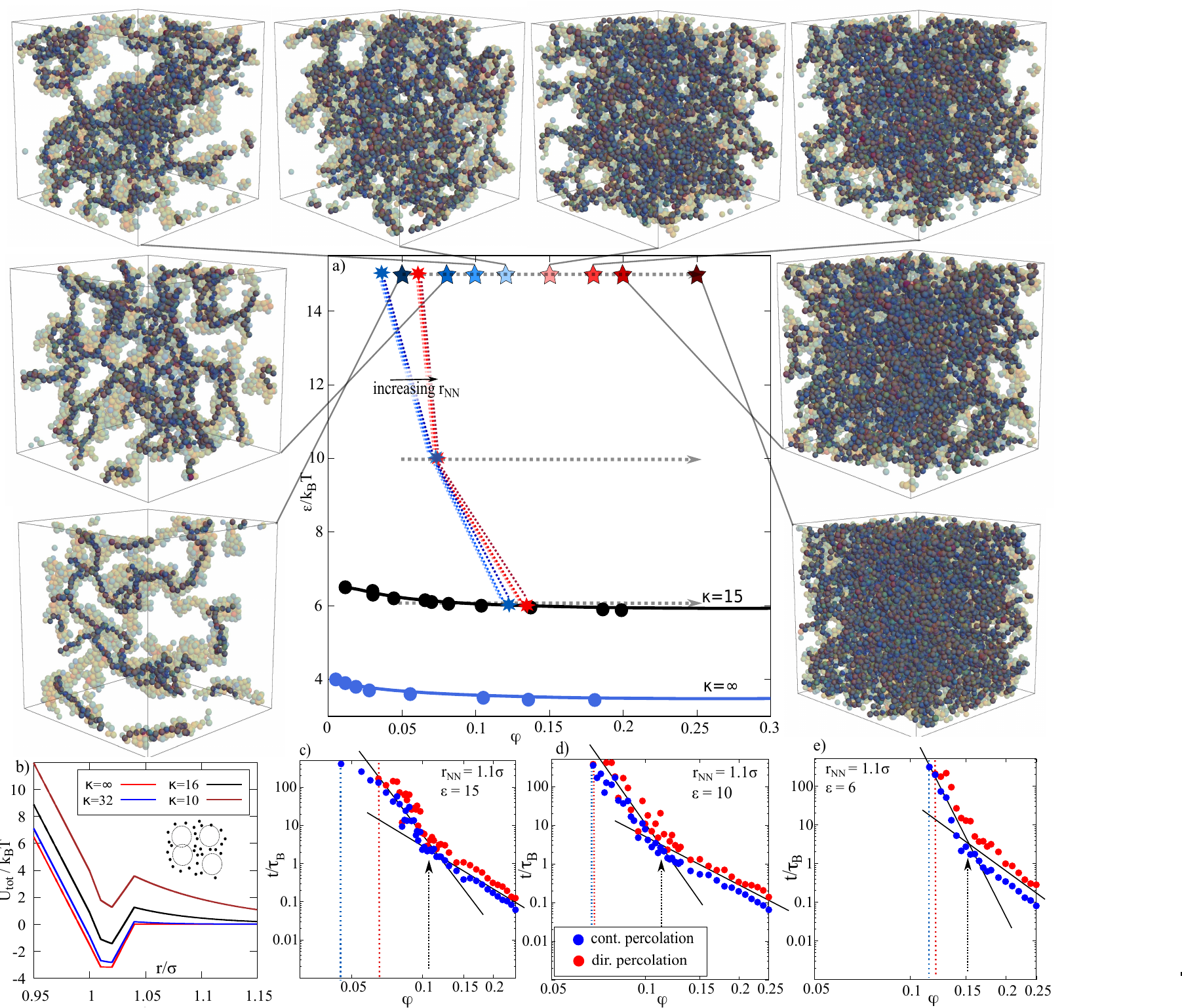}
\caption{\label{fig:phaseDiagram} \textbf{a)}Phase diagram of our gel forming model system. The insets show examples of gel networks at different packing fractions far above the binodal line for $\epsilon=15$ after simulation time $2000\tau_B$. The opaque particles represent the complete network, while the non-opaque particles form the reduced network obtained by our backbone method \cite{gimperlein}. The grey dotted arrows indicate lines along which we have simulated systems for analysis in this article.
  \textbf{b)} shows the interaction potential for $\epsilon=3$ and different values of $\kappa$. \textbf{c)-e)} show the percolation time for systems simulated along the grey opaque arrows in a) shown above for a next neighbor distance of $1.1\sigma$ (\textbf{c)} $\epsilon=15$, \textbf{d)} $\epsilon$=10, \textbf{e)} $\epsilon$=6). The red dotted line corresponds to directed percolation threshold, while the blue dotted line represent continuous percolation threshold. For all values of $\epsilon$ the slope in the double logarithmic plots changes somewhere between  $\varphi=0.1$ and $\varphi=0.15$. Systems for p[ercolation analysis were simulated for at most $500\tau_B$ and the smallest at least once percolating density was taken as the percolation threshold. The estimates for percolation transition lines are drawn as dashed lines in the phase diagram and depend on the chosen next neighbor distance.} 
\end{figure*}

\subsection{Colloid-Polymer mixture}
We study a gel forming model system that is similar to the system considered in \cite{speck2} and that we have previously used in \cite{gimperlein}.
A colloid-polymer mixture is simulated with Brownian dynamics simulations. The colloids possess a polydispersity of $5\%$ and the effective colloid-colloid interactions are modeled by summing a short-ranged square-well-potential $U_{SW}(r)$ and a longer ranged repulsive Yukawa tail $U_{YK}(r)$, such that $U_{tot}(r)=U_{SW}(r)+U_{YK}(r)$, where
\scalebox{.8}{\parbox{\columnwidth}{
\begin{align*}
U_{\text{YK}, ij}(r)&=C \left( \frac{2}{2+\kappa
\sigma_{ij}}\right)^2\left(\frac{\sigma_{ij}}{r}\right)\text{exp}[-\kappa(r-\sigma_{ij})],\\
U_{\text{SW},ij}(r)&=\begin{cases}
-\frac{\epsilon}{2\alpha_{ij}} r+\frac{\epsilon(\sigma_{ij}-\alpha_{ij})}{2\alpha_{ij}} & r<\sigma_{ij}+\alpha_{ij}\\
-\epsilon & \sigma_{ij}+\alpha_{ij} \leq r \leq
\sigma_{ij}+\delta_{ij}-\alpha_{ij}\\
\frac{\epsilon}{2\alpha_{ij}}r - \frac{\epsilon(\sigma_{ij}+\delta_{ij}+\alpha_{ij})}{2\alpha_{ij}}
& \sigma_{ij}+\delta_{ij}-\alpha_{ij} < r < \sigma_{ij}+\delta_{ij}+\alpha_{ij} \\
0 & \text{else}.
\end{cases}
\end{align*}}}
A sketch of the potential is shown in Fig. \ref{fig:phaseDiagram} b). Here $\sigma_{ij}=r_i+r_j$, where $r_i$ is the radius of particle $i$. The strength of the attraction is modeled by the parameter $\epsilon$, which is the depth of the square well-potential. The range of attractive interaction is given by the width of the square well, i.e.,  $\delta_{ij}=0.03\sigma$. To avoid infinite forces, we smoothen the square-well potential by the additional parameter $\alpha_{ij}=\frac{1}{5}\delta_{ij}$. In experimental setups the screening length $\kappa^{-1}$ can be tuned by
modifying salt concentration \cite{kohl}. In our theoretical setup it is responsible for the strength and range of the repulsive force. For $\kappa=\infty$ the Yukawa-tail vanishes and therefore this
represents the case of a purely attractive potential. The parameter $C$ is chosen as 200$k_BT$. For our simulations we choose a cut-off distance at $\frac{r_{\text{Cut}}}{\sigma}=1+\frac{4}{\kappa}$ as in
\cite{speck2}. In summary the whole system can be characterized by choosing a triplet of parameters $(\epsilon, \kappa, \varphi)$, where
$\varphi=\frac{\pi}{6}\sigma^3\frac{N}{L^3}$ is the packing fraction of the
system with the box size $L$. Note that in the following $\epsilon$ is
used in units of $k_BT$ and $\kappa$ in units of $\sigma^{-1}$.

\subsection{Brownian dynamics simulations}
In our simulations the motion of collidal particles is determined by the overdamped Langevin-equation for particle $j$
\begin{align*}
\gamma \frac{d}{dt}\vec{r}_j=\vec{F}_{\text{int}}+\vec{F}_{\text{th}},
\end{align*}
which is numerically integrated. $\gamma$ is the friction constant and $\vec{F}_{\text{int}}$ models
the effective force between colloidal particles as given by the pair
interaction potential introduced in the previous subsection. Thermal fluctuations induce random motion of particles which have to obey two conditions, such that the correct diffusion behavior is reproduced. These random forces are denoted by
$\vec{F}_{\text{th}}$ and fulfill $\left\langle \vec{F}_{\text{th}}\right\rangle=\vec{0}$ and
$\left\langle F_{\text{th},i}(t)F_{\text{th},j}(t')\right\rangle=2\gamma
k_\textnormal{B} T\delta_{ij}\delta(t-t')$. Forces are calculated using a parallelized version of a combination of the Verlet-list algorithm and the linked-cell algorithm to minimize computation time.\cite{allen}  The time constant in our simulations is given by the Brownian time $\tau_B=\frac{\sigma^2\gamma}{4k_\textnormal{B} T}$ and we use a time resolution of $\Delta t=10^{-5} \tau_B$. We use periodic boundary conditions and boxes of size $30\sigma\times 30\sigma\times 30\sigma$, where $\sigma$ is the mean particle diameter, for all considered packing fractions. The boxes are then randomly filled with particles until the considered packing fraction is reached. Due to the polydispersity of the system the number of particles for a given packing fraction may vary slightly.

\subsection{Continuous and directed Percolation}

Fig. \ref{fig:phaseDiagram} shows the phase diagram of the investigated system. The binodal line in the phase diagram was obtained by extracting coexistence densities from several simulations. The additional dotted lines in blue and red show the percolation behavior of the system.
The determination of the percolation behavior was done using a next neighbor distance of $1.036\sigma$, $1.1\sigma$ and $1.2\sigma$ and depends on the choice of this distance. We have chosen the bigger distances to account for influence of thermal fluctuations of the particles and make the results more stable. We distinguish continuous and directed percolation. In both cases system spanning paths exists, but in the case of continuous percolation steps in all directions - including backward steps - are allowed, while in directed percolation only steps in a previously chosen arbitrary direction are considered.\cite{kohl} The percolation time is defined as the time until the first continuous or directed percolating path in the network is found. This time is plotted for different attraction strengths as a function of packing fraction in Fig. \ref{fig:phaseDiagram}c)-e). The dotted lines in these plots show an estimation to the percolation threshold for the given attraction strength. Simulations were carried out for at most 500$\tau_B$. The simulation time also influences the exact position of the percolation lines in the phase diagram. As a compromise of computational effort and equilibration of the system, we have chosen $500\tau_B$ as the maximum simulation time. The different results for the different choices of neighbor distances are represented by the different lines in the phase diagram in Fig. \ref{fig:phaseDiagram} and show that the exact position of the percolation threshold depends on these choices though the differences are small. Also, as gels are non-equilibrium systems they still undergo slow dynamical changes and as a consequence the exact value of the percolation threshold might depend on the simulation time. As the relaxation dynamics slows down at the directed percolation line \cite{kohl}, the waiting time dependence of this transition is most pronounced, especially for simulations of dilute systems and at low values of $\epsilon$, as here strands of particles can split up and connect easily. As the value of the percolation time we have taken the first time step at which we could find percolation irrespectively of how stable or unstable the percolating strands are. Note that continously percolating strands often are significantly more stable than strands with directed percolation. That means that at least for dilute networks at lower attraction strengths the directed percolation connections were destroyed often in less than 1$\tau_B$ and eventually reformed again. Thus constraints due to percolation are most important deep inside the phase separated region but not close to the binodal line (or even for attractions smaller than at the binodal line).

\subsection{Methods and skeletonization process}

\begin{figure}[t]
\includegraphics[width=\columnwidth]{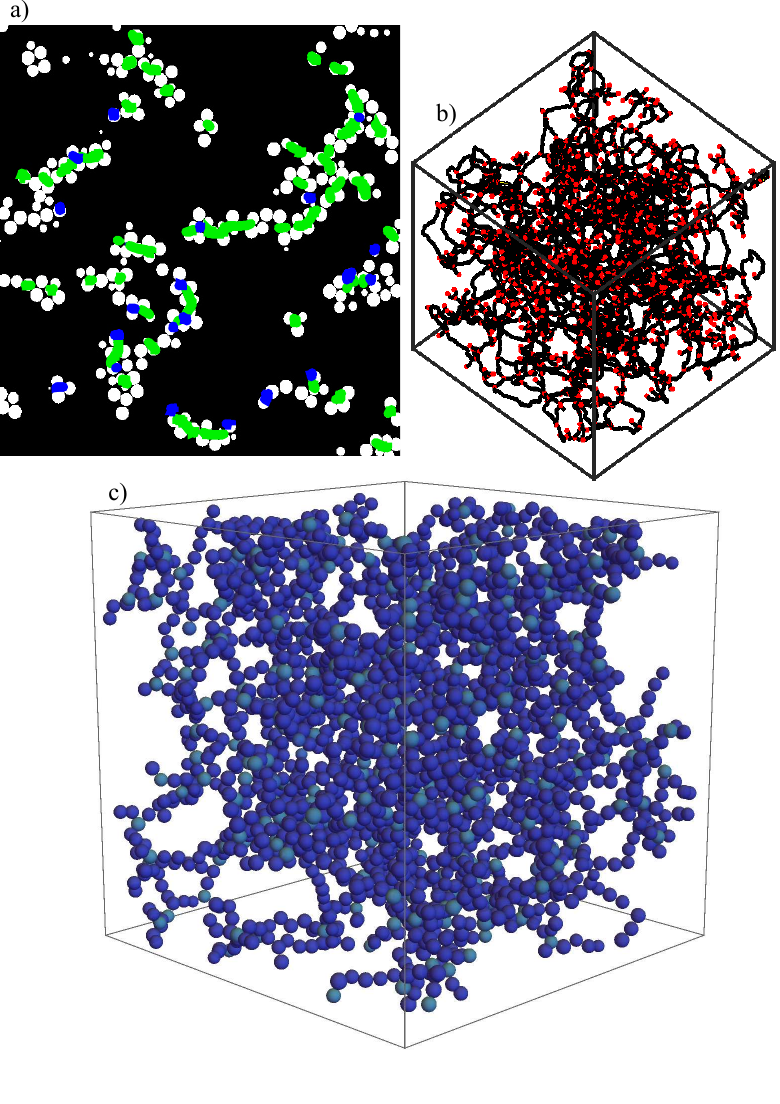}
\caption{\label{fig:methodsSkeleton}
Reduced networks obtained for a gel structure with packing fraction $\varphi=0.18$ at time $2000\tau_B$ obtained \textbf{a, b)} with ArGSLab\cite{immink} and \textbf{c)} with our backbone method introduced in \cite{gimperlein}. While ArGSLab leads to a skeleton network of nodes and links (in the slice shown in \textbf{a)} nodes are shown in blue and the links in green), our backbone method leads to a network of colloids that is obtained if all particles not needed for the essential connections are left away. Both methods are used in this article depending on the property that we want to determine.}
\end{figure}

In our global structural analysis of gel networks we use two different skeletonization methods. The first method is on a particle level and described in detail in our last article \cite{gimperlein}. It is based on the idea of preserving only crucial connections in the network. Therefore, we delete colloids as long as the connectivity of the whole network is not destroyed. Like this we get a backbone of the original network on a particle level. Examples of our algorithm for different packing fractions are shown in the phase diagram in Fig. \ref{fig:phaseDiagram}. The computation of reduced networks in this way is computationally expensive, therefore we also use ArGSLab to calculate skeleton structures of gel networks.\cite{immink} ArGSLab is a software designed for quantitative comparison of experimental and computational data of particle networks and determination of reduced structures. It is based on binarization of the original input and a thinning algorithm, explained in more detail in the original publication.\cite{immink} The advantage is that it performs much faster than our reduction algorithm, its disadvantage is that it is pixel and not particle based. Furthermore, it is mainly designed for experimental data and therefore does not take the periodic boundary conditions into account that we use in our simulations. Results obtained with the two methods are shown in Fig. \ref{fig:methodsSkeleton}. %One of our goals is to see how the two different methods compare to each other and see in which case either one of the two should be used. WO VERGLEICHEN WIR DIES?
One of our goals is to show how these two inherently different methods can profit from each other and can be used together to get insight into different structural properties of complex network structures. The suitable type of skeletonization algorithm depends on the task which is of interest.

The skeletons obtained from one of the two methods are used to calculate different distribution functions that we introduce in the following. We call two particles $i$ and $j$ neighbors, if their distance $r_{ij}\leq 1.036\sigma$, i.e., if it is smaller than the width of the attractive step in the interaction potential. After the skeletonization with our analytic approach we extract distribution functions.

For the loop size and the link length distribution we interpret the particle network as a graph structure by identifying particles with nodes and connecting two nodes by an edge if the corresponding particle distance is smaller than the next neighbor distance. For loop size distribution we then determine the minimal cycle basis of the graph structure, the weights on the edges are given by the particle distances. The minimal cycle basis is a set of loops with minimal weight, from which all loops in the network can be reconstructed. For link length distributions we first determine all crossing points or terminal points in the network. These are found by neglecting all particles with next neighbor count two, because these have to be particles in the middle of strings. Then the length of the shortest path from each crossing or terminal point to the next neighboring crossing or terminal points is calculated and interpreted as the link length.

The pore size distribution is obtained by choosing random points in the system and calculating the minimal distance from this point to any other point in the particle network. This is the size of the biggest sphere centered at the chosen point, not overlapping any particle in the system. \cite{kob,torquato1,torquato2,torquato3,torquato4}

The last property we consider in this context is the link thickness. This is of course extracted not from skeletonized networks, but from the entire network structure by generalizing a 2-dimensional approach from \cite{dauchot} to 3-dimensional systems. We insert random planes in the system and calculate which particles intersect the plane. From the intersecting particles we detrmine connected components using the graph structure and itnerpret the size of the connected components as the string thickness. When interpreting the results from this calculation one has to consider that the planes do not have to be perpendicular to the strings in the network, therefore the tail of the distribution does also show an estimate for the longest straight connection.

\section{Results}
\label{sec3}

The phase behavior of the system has been discussed previously in \cite{gimperlein}. The phase diagram is shown in Fig. \ref{fig:phaseDiagram}. 
As stated in the introduction a rich phase behavior can be observed. The crossover between the dilute and clumpy gel network case is not sharp but rather continuous and smooth and will be further analyzed in this article.

\subsection{Analysis of reduced networks}
\label{sec3.1}

\subsubsection*{Phase diagram and percolation times} We first evaluated the percolation time of the networks as a function of packing fraction.
As one would expect, the percolation time is decreasing with increasing packing fraction. For high densities the network percolates almost immediately. Interestingly in the double-logarithmic plots in Figs. \ref{fig:phaseDiagram} \textbf{c)}-\textbf{d)} we see two different regimes and a crossover at a packing fraction between $\varphi\approx 0.1$ and $\varphi\approx 0.15$, depending slightly on the value of $\epsilon$. In both the low and the high density regime the percolation time decreases with increasing packing fraction. The different slopes in the two regimes may be a hint to the different mechanism behind gelation. However, directly at the binodal line $\epsilon=6$ we do not see different regimes concerning the percolation behavior.
As for larger attractions, the percolation behavior changes, we also want to find out which structural properties might change. To investigate this we use ArGSLab and our backbone approach to analyze the gel structures.

\begin{figure}[t]
\includegraphics[width=\columnwidth]{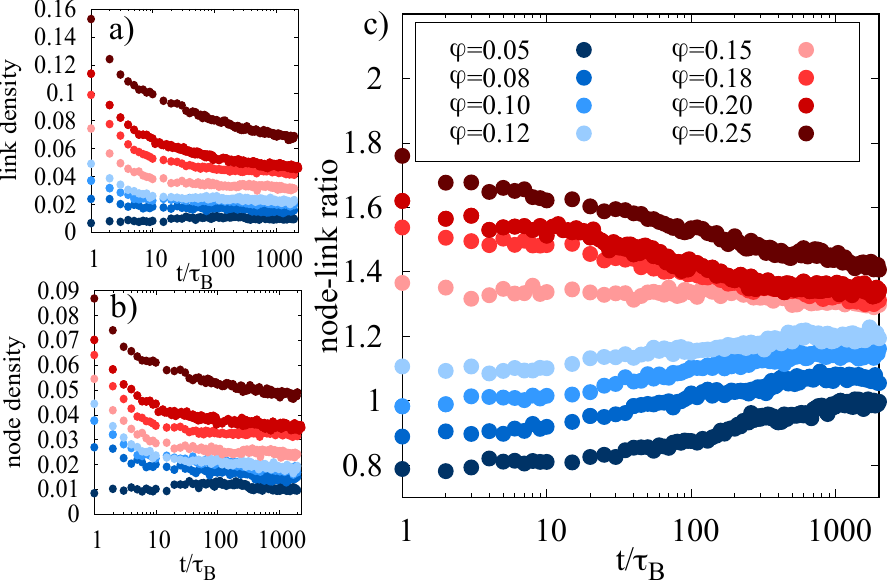}
\caption{\label{fig:reduction}\textbf{a)},\textbf{b)} Evolution of node and link density calculated as the number of nodes and links after the skeletonization divided by the volume of the simulation box. \textbf{c)} Ratio of nodes and links in the skeletons. In dense gels a constant value is approached from above, while in dilute networks the approach is from below. The data is taken from skeletons obtained with ArGSLab.}
\end{figure}

\subsubsection*{Nodes and links in the reduced network} The dynamical evolution of the backbone structure is analyzed using ArGSLab. At first we compare the number of nodes and links which remain in the network after applying the skeletonization algorithm. As expected the number of remaining links and nodes decreases as a function of time for all considered packing fractions as the structure of the gel coarsens and therefore more links and particles get neglectable (see Figs. \ref{fig:reduction}\textbf{b)} and \textbf{c)}). Therefore, during the dynamical evolution of the gel particles form bigger clusters for all considered packing fractions. Interestingly if we plot the ratio of links and nodes (see Fig. \ref{fig:reduction}\textbf{a)}) we see a difference between dilute networks and dense networks. While in dilute networks a constant ratio is approached from below, in dense gels a constant value is approached from above. Therefore, ageing in dilute gel networks seem to occur by thinning out the network even further and especially getting rid of links. In contrast, ageing in clumpy gels is related to growing clusters such that the number of nodes is reduced faster than the number of links. The value for the node to link ratio that is approached depends on the packing fraction. Interestingly the differences between dilute networks and clumpy gels is much larger at small times but decreases during ageing. The remaining differences between dilute and dense systems seem to be stable over a long time. However, we cannot rule out the possibility that the node to link ratio might not be constant for very long times but that there maybe is a further approach due to ageing on much longer timescales. The change of the dynamics occurs around $\varphi=0.12-0.15$. This value coincides with the dynamical percolation crossover point mentioned earlier and can be seen as the result of different dynamical pathways in gelation.

\begin{figure}[t]
\includegraphics[width=\columnwidth]{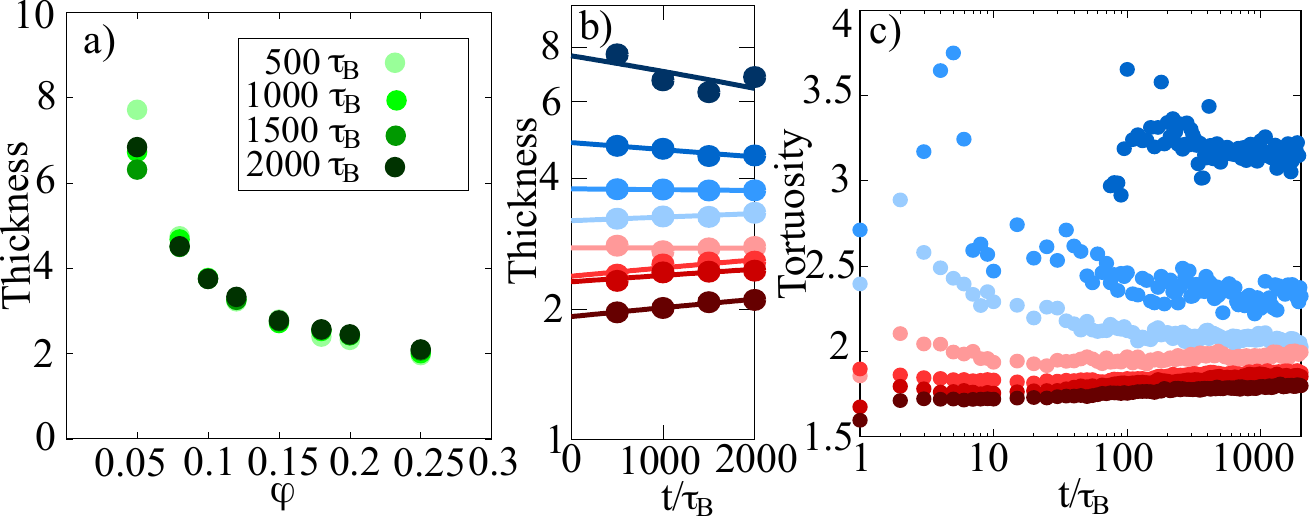}
\caption{\label{fig:tortuosity}
\textbf{a)} Thickness of strings in the network as a function of packing fraction at different simulation times.  In general strings tend to get thinner with increasing packing fraction. \textbf{b)} Thickness of strings as a function of time. Fitted linear functions are show to illustrate the general trend. As a function of time the thickness decreases for dilute networks, but increases slightly for dense gels. \textbf{c)} Tortuosity of gel networks. Dense networks tend to form rather straight connections, while dilute networks are more bent. The colors in \textbf{b)} and \textbf{c)} are the same as in Fig. \ref{fig:reduction}\textbf{c)}.}
\end{figure}

\subsubsection*{Thickness and tortuosity of strands} 
Using our analytic backbone construction method, we are able to get skeletons on a particle base. This can be used to get insight into the structure of the single connection strings in the network. To analyze how the thickness of these strings in the network changes as a function of packing fraction we calculate the quantity
\begin{align*}
t(\varphi)=\frac{N_{\text{tot}}(\varphi)}{L_{\text{skel}}(\varphi)}
\end{align*}
where $t(\varphi)$ stands for the thickness of the strings as a function of packing fraction and is calculated by dividing the total number of particles in the non-skeletonized network $N_{\text{tot}}(\varphi)$ by the total length of the skeletonized network $L_{\text{skel}}(\varphi)$.  The length of the skeletonized network can be calculated by summing up the diameters of all particles in the backbone structure. 
This gives a measure of how many particles are in average found per particle diameter in the non-skeletonized network and can be seen as an approximation to the thickness of the strings. We find that the thickness decreases with increasing packing fraction. This shows that in dilute networks particles tend to cluster more easily, while in dense networks the small number of particles and their pairwise interaction forces them to form thinner strings. Another observation we show in Fig. \ref{fig:tortuosity}b) is that in dilute networks the strings get thinner with time, while in dense networks they tend to get thicker. The evolution towards thicker strings in dense networks is nevertheless quite slow. Again the crossover point seems to be at approximately $\varphi=0.1-0.12$. This may again be an indicator of different dynamical pathways in gelation. Dilute gel systems contain random clusters in the beginning of the evolution. Then strands are formed that initially are loose and might be clumpy. As additional particles may move quite freely and strands can easily reorganize, the evolution leads to thinner and less bent strands in the network. In contrast in dense gels there are more conditions concerning the movement of particles and therefore the confined movement leads to the early formation of strings. In other words, the initial random clumps are not compact clusters, but networks consisting of thin strands. The process to become more compact is very slow.\\
We also show the tortuosity of the gel structures in Fig. \ref{fig:tortuosity}\textbf{c)}, which is a measure of the bending of connections between nodes. The length of percolating paths between nodes at the upper and lower boundary of the box is calculated and divided by the euclidean distance of the nodes, which represents the length of a straight connection line. Therefore toruosity can be interpreted as a fraction of how bent a connection is compared to a straight line. Higher values stand for more bent lines, while values close to 1 are observed for straight connections. A careful analysis shows that erraticity of strands plays a role for tortuosity, but also the number of pathways/branching nodes is an important factor. We see that dilute networks are in general more bent than dense networks and that the tortuosity of dilute networks changes much more drastical during gelation, especially in the initial 10-20$\tau_B$. According to the analysis with ArGSLab, the network at $\varphi=0.08$ starts to percolate after approximately 100$\tau_B$. Therefore, prior to that time no data points can be calculated. Note that the tortuosity for very dilute packings like $\varphi=0.05$ cannot be calculated using ArGSLab, as the analyzed network never percolates (except if periodic boundary conditions were taken into account, which ArGSLab does not do).
The tortuosity of dense gels changes very little over time. It is even slightly increasing during temporal evolution, while the tortuosity of dilute networks decreases until it reaches a constant value. A general trend is visible in all data sets: The higher the packing fraction, the lower the tortuosity.
\begin{figure}[t]
\includegraphics[width=\columnwidth]{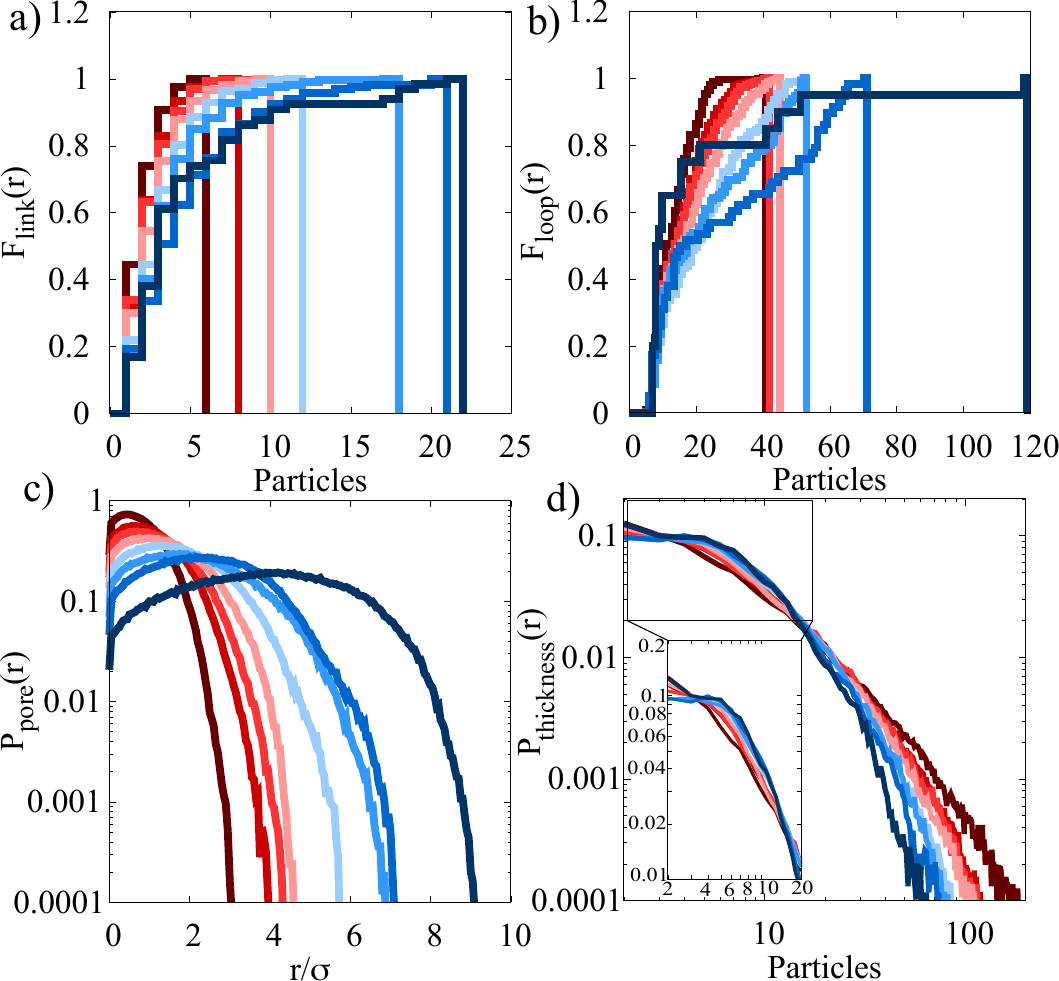}
\caption{\label{fig:distributions} \textbf{a)} Cumulative distribution of link length in reduced networks of packing fraction $\varphi=0.05-0.25$. \textbf{b)} Cumulative distribution of loop sizes in reduced networks of packing fraction $\varphi=0.05-0.25$. \textbf{c)} Distribution of pore sizes in reduced networks of packing fraction $\varphi=0.05-0.25$. \textbf{d)} Distribution of string thinckness in non-reduced networks. The general trend can be seen in all of the distributions: The higher the density of the system, the shorter the links between nodes and the smaller the loops and pores in the network. The analysis of the string thickness confirms the result that individual strings in dilute networks are thicker than in dense networks, but dense networks show longer straight strings. All figures show distributions obtained from simulations after $2000\tau_B$ and the colors are the same as in Fig. \ref{fig:reduction}\textbf{c)}.}
\end{figure}

\subsubsection*{Structural distribution functions} After looking at the temporal evolution of the strings themselves, we focus on the global structure of the whole system. Therefore we look at loop size, link length, pore size\cite{kob} and string thickness distributions. These are calculated as described in the methods section in Sec. \ref{sec2}. 
In Fig. \ref{fig:distributions} loop size and link length are shown as cumulated distributions, while pore size and link thickness are shown as a non-cumulative distribution. We observe the general trend that increasing the density leads to a decrease in loop size, link length, pore size, and also link thickness which means that the network itself gets more compact and the void spaces in between get smaller. Especially the decrease in loop size is very pronounced. The biggest loop for a network at $\varphi=0.05$ is nearly twice as big as the biggest loop for the network with $\varphi=0.08$ even if the difference in packing fraction is quite small. The distribution of the link thickness in Fig. \ref{fig:distributions}d) reveals two interesting properties. The first one can be read of from the behavior at small particle numbers - here the dilute systems show bigger values as the dense systems. This can be taken as a measure for the link thickness and agrees with our result from Fig. \ref{fig:tortuosity}a). The second interesting property is taken from the tail of the distribution. As the planes, which are used to calculate the distribution functions are chosen arbitrarily, it can happen that they do not intersect a string perpendicularly, but are rather parallel to the string direction. Therefore the tail of the distribution can be seen as an approximation to the longest straight strings in the system. The result that dense network have longer, straight links is in accordance with our result from the tortuosity calculation in \ref{fig:tortuosity}b) where we find smaller tortuosity values for dense systems.

\begin{figure}[t]
\includegraphics[width=\columnwidth]{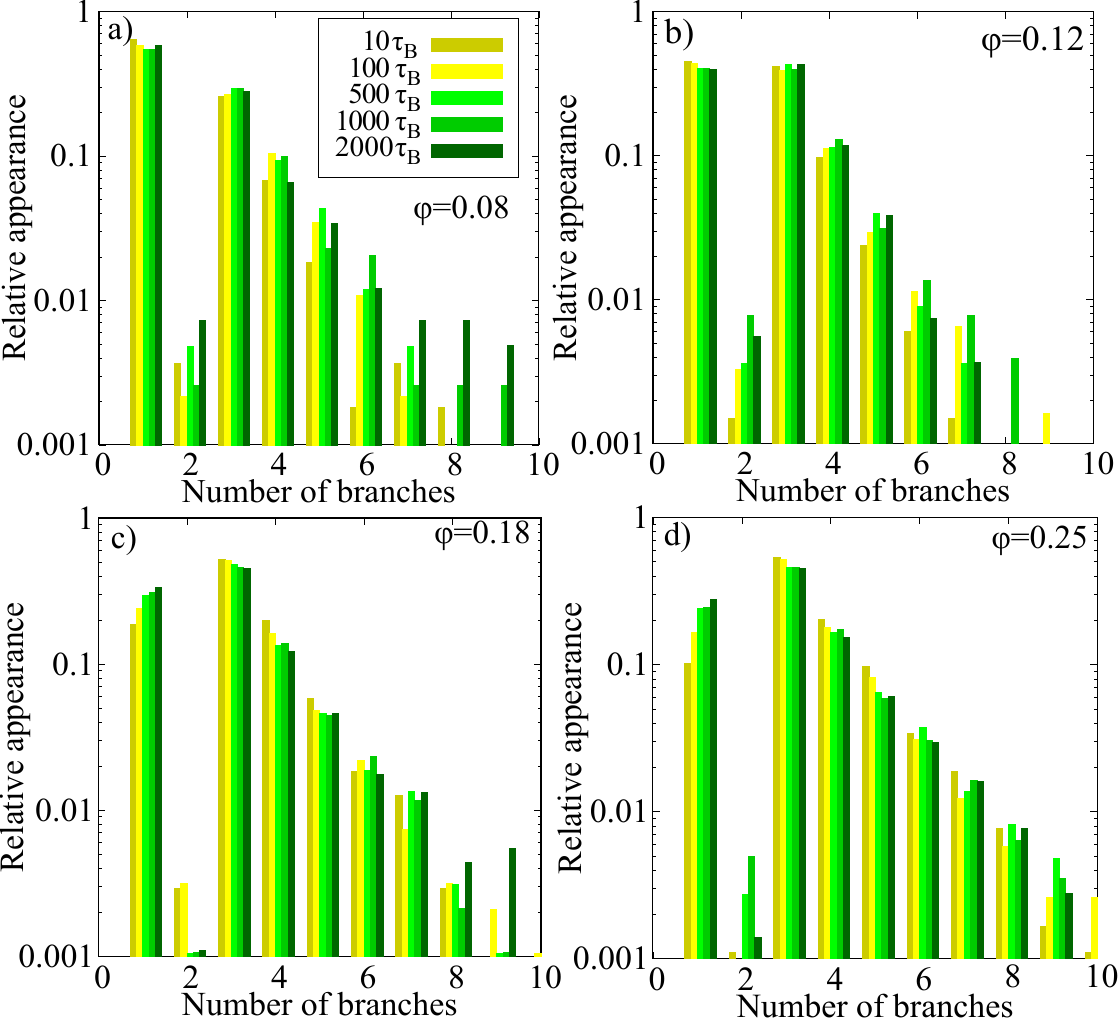}
\caption{\label{fig:branches} Relative number of branches connected to each node in the reduced network as calculated using ArGSLab. Different colors indicate different timesteps.}
\end{figure}

Concerning the number of branches which are connected to a single node in the graph we use ArGSLab to calculate distributions for different packing fractions. In Fig. \ref{fig:branches} the relative appearance of nodes with the given number of branches is shown. Nodes with one branch are therefore terminal nodes of strings and no crossing points. Nodes with two branches are strongly suppressed and in a further idealized network should be exactly zero as these nodes correspond to particles inside a link, i.e., the outgoing links could be replaced by one link and the node could be left away. But depending on the used cleaning routines in the algorithm small branches can still contribute and non-zero values can arise.

For nodes with more than two branches we see that the relative appearance decreasess exponentially as a function of branch number. But if the temporal evolution is considered we can again see a difference between dilute and dense systems. For dilute systems the relative appearance of nodes with more branches increases with time, while for dense systems it decreases slightly or stays constant. This is possible as the fraction of terminal nodes decreases for dilute systems and increases for dense systems with time. Therefore, while free ends and loose strands are reduced in dilute networks, in dense gels the surface of growing clusters increases and thus the number of terminal nodes.

\subsection{Occurence of local ordering}
\label{sec3.2}

Local ordering plays a key role in understanding the way towards gelation and dynamical arrest. Recently it was reported that experiments of dilute, stress-free gel networks show that local ordering happens in an hierarchical manner and the different stages of the ordering process are responsible for distinct properties of gels.\cite{tanaka} Stress-free gel networks means that the packing fraction is lower than $\varphi=0.1$ where percolation only occurs after the gelation. Using our simulation results, we want to check the results obtained for dilute gel networks and in addition extend the range of studied packing fractions to denser networks.
In this article we extend the range of studied packing fractions up to $\varphi=0.25$ and also look at differences between weakly or strongly bound particle networks.

As the most important local ordering structures which occur we consider Tetrahedra, 2-Tetrahedra, 3-Tetrahedra and Pentagonal bipyramides (PBP). These structures and especially the fractions of particles which are part of these structures are detected using the same analysis methods as in \cite{tanaka}. It is important to note that a particle can be part of several local orderings at the same time, therefore the distribution curves do not add up to 1. During the analysis two particles $i$ and $j$ are seen as neighbors if their distance $r_{ij}<1.1\sigma$. This is in contrast to our previous definition of neighborhood from the first part of the article where we choose $r_{ij}\leq 1.036\sigma$. This difference is made to account for thermal fluctuations and ensure more stability in the calculations to reduce fluctuations of neighborhood relations during the analysis.

\begin{figure}[t]
\includegraphics[width=\columnwidth]{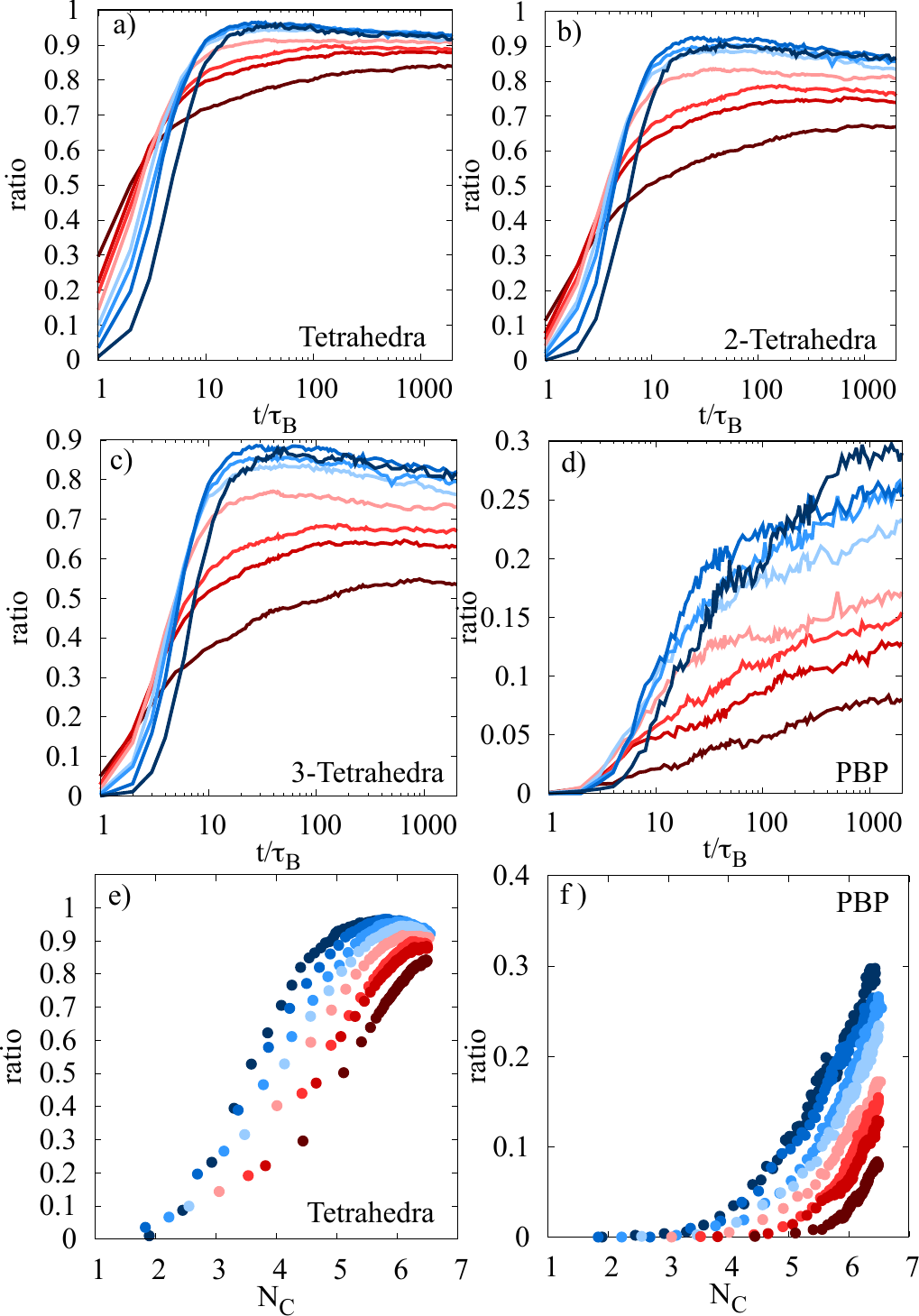}
\caption{\label{fig:localstructures1} Fraction of particles inside distinct local structures. The hierarchical formation of local structures becomes apparent for all kinds of local structures analyzed. The packing fraction influences the fraction of particles in certain local stuctures and also the speed of local structure formation. \textbf{e)} Particles in Tetrahedral clusters as function of coordination number. The point of the maximum shifts with increasing packing fraction to higher values of $\langle N_c\rangle$. For dilute networks the maximum is reached before $\langle N_c\rangle=6$. \textbf{f)} The fraction of particles in PBP-clusters is directly proportional to $\langle N_c\rangle$, but the onset of the formation of PBP-structures shifts to higher values of $\langle N_c\rangle$ for increasing packing fractions.}
\end{figure}

The hierarchical formation of local ordering mentioned in \cite{tanaka} for dilute gel networks is reproduced not only for dilute networks, but also for dense gels. Initally particles form tetrahedral clusters, which merge to 2-tetrahedral clusters, 3-tetrahedral clusters and finally to pentagonal bipyramidal clusters. We see that for dilute gel networks the fraction of particles inside local structures is higher then for dense networks, indicating higher local order in dilute networks. Especially for the pentagonal bipyramides the fraction of particles for the system with $\varphi=0.25$ is only $10\%$ after $2000\tau_B$, while for the dilute network with $\varphi=0.05$ it is nearly $30\%$. A difference which can be seen between dilute and dense networks is that for all dilute networks ($\varphi \leq 0.12$) the ratio of particles in local structures is approximately the same, while for dense networks the final ratio depends on the packing fraction. Concerning the dynamics of local structure formation we see that for dense systems the initial formation is fast, but later slows down, and is ultimatively overtaken by the dilute systems. This may indicate differences in the evolution mechanism. Dense systems percolate quickly as already stated in sec.\ref{sec2}. Therefore, mechanical stress due to percolation influences the evolution of the system and makes it more difficult to form new local structures (or to form local structures at all). This explains both the slower dynamics after rapid initial evolution and the lower fraction of particles in later stages of the hierarchical structure formation chain. Interestingly the fraction of particles in all of the tetrahedral states reaches a maximum and decreses afterwards, which means that tetrahedral order constitutes an intermediate stage during gelation and may slowly vanish during the further evolution of the system. In \cite{tanaka} it is also stated that the formation of pentagonal bipyramides is connected to the onset of solidity, as the fraction of particles in PBP-clusters is directly proportional to the mean coordination number at the isostatic point $\langle N_c\rangle=6$. This behavior is reproduced for low packing fractions and also observed for higher packing fractions. In contrast, in dense gels the fraction of particles in tetrahedral clusters is directly proportional to the mean coordination number at $\langle N_c\rangle=6$. This means that for dense gels the formation of tetrahedral clusters may also play a role for the onset of solidity. Interestingly, for lower packing fractions the tetrahedral curves show a maximum at values $\langle N_c\rangle<6$, which means that for higher coordination numbers the fraction of tetrahedral clusters decreases. For larger packing fractions this maximum shifts to larger values of $\langle N_c\rangle$ and disappears for the largest considered packing fractions.

\begin{figure}[t]
\includegraphics[width=\columnwidth]{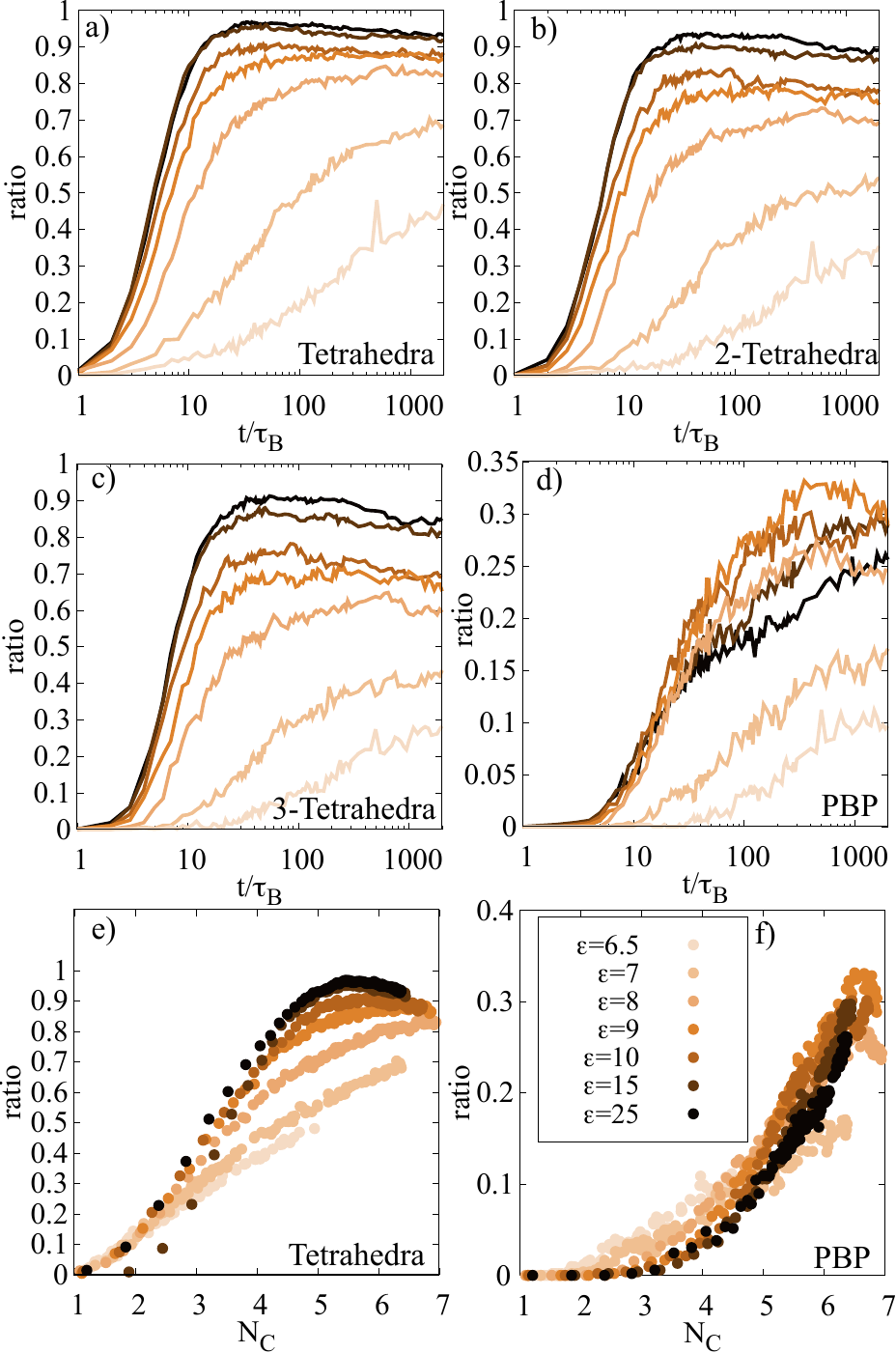}
\caption{\label{fig:localstructures2} Fraction of particles inseide distinct local structures for $\varphi=0.05$ as a function of attraction strength $\epsilon$. For the tetrahedral-structures in \textbf{a)}-\textbf{c)} the fraction of particles inside the structure decreases with decreasing attraction $\epsilon$. For the PBP-structures the particle fraction is similar for all network states ($\epsilon>7$) while it is lower for the cluster states ($\epsilon \leq 7$). Fraction of particles in tetrahedral structures and in PBP-structures as a function of the mean coordination number. For PBP structures all cluster fluids show similar behavior and all dilute gel networks show similar behavior, irrespective of the attractive strength used. So this behavior seems to be a general feature of the studied structural regime not depending on the exact attractive strength $\epsilon$.}
\end{figure}

Finally, we study how the formation of local structures depends on the interaction between the particles by changing the strength of the  attractive forces. This means we analyze how homogeneous fluids, cluster fluids and dilute gel networks differ structurally at identical packing fraction. Therefore a system at $\varphi=0.05$ is studied at different values of $\epsilon$ for $2000\tau_B$. As expected for homogeneous fluids, for a system simulated below the binodal and spinodal line at $\epsilon=4.5$ no structure formation at all is seen. As the phase separated regime of the phase diagram is entered we see structure formation as well in cluster fluids (small values of $\epsilon$) as also in dilute gel networks (higher values of $\epsilon$). We see a similar trend for all of the tetrahedral structures: The higher the attraction between the particles, the faster these local structures are formed and the higher the fraction of particles in these structures is. For the PBP-structure the trend seems to be different. Fig. \ref{fig:localstructures2}d) shows that for all the network states ($\epsilon>7$) the fraction of particles in PBP-structures approaches the same value of about $0.25-0.3$ in approximately the same time. In contrast, the cluster fluid states ($\epsilon=6.5$ and $\epsilon=7$) show a lower fraction of particles in PBP-structure. This means that the gelation and chain formation seems to be a consequence of the PBP-structure formation. The less clumpy
%more "network-like"
a system is, the higher the fraction of pentagonal bipyramides. We again analyze the dependence of structural ordering on the mean coordination number of the system in Fig. \ref{fig:localstructures2}e) and f). For the system with $\epsilon=6.5$ the solidity threshold $\langle N_C \rangle=6$ is never reached. We notice that for cluster fluids the pentagonal ordering is already present at very small coordination number and increases linearly, while for the network states the pentagonal ordering becomes important at higher coordination numbers. This behavior seems to be more or less independent of the attractive strength, depending only on the regime (cluster fluid or network state) in the phase diagram which is studied, similar to the results for the PBP-fraction in Fig. \ref{fig:localstructures2}d).

\section{Conclusions}
\label{sec4}
Our simulations and analysis of gel networks at different packing fractions and intraparticle attractions reveal different pathways in gelation influencing dynamical properties on the one hand and structural properties of the emerging networks on the other hand.

Using two different skeletonization algorithms we find that on a global scale at a packing fraction of approximately $\varphi \approx 0.12$ the gelation dynamics changes. This is related to early percolation dynamics in the system. In Fig. \ref{fig:phaseDiagram}c)-e) we show the percolation time of the system as a function of packing fraction. Between $\varphi\approx 0.1$ and $\varphi\approx 0.15$ (depending on the chosen attractivity $\epsilon$) the slope of the double logarithmic plot changes. This is an indicator of different dynamical regimes. For dense systems the percolation time is very small ($\leq 1\tau_B$) leading to relaxation which is restricted by mechanical stress. In contrast, for dilute networks the system can first form disconnected clusters which can evolve freely and start to percolate later.

This has consequences for different properties of the gel. We find that the thickness of the strings decreases with increasing packing fraction. For dilute gel networks the string thickness decreases with time, while for dense gels it increases. The crossover packing fraction is again around $\varphi\approx 0.12$. %Still one has to mention that at least for the dense networks this effect is small at the times we analyzed.
Note that the overall structure of the gel networks gets more compact with increasing packing fraction, in the sense that loop size, link length, pore size and string thickness all decrease with increasing packing fraction.

Regarding the influence of the packing fraction on the local structure formation we have analyzed the fraction of particles which are located inside different types of tetrahedral or pentagonal substructures. We find that for dilute gel networks $\varphi<0.15$ the final fraction of particles in tetrahedral subclusters is more or less independant of the packinbg fraction, all of our systems approach similar values. Concerning the dynamics of structure formation in dilute gel networks one notices that the more dilute the network is, the slower the initial structure formation, but once initial structures have formed the curves are more or less parallel to each other. For dense systems we see a different behavior. Here we notice that the final fraction of particles in all of the mentioned local structures decreases with increasing packing fraction, which leads us to the assumption that dense networks show a weaker local order.

Turning away from the effect of packing fraction, we have also investigated the effect of attraction strength. Here we notice, as might be expected, that higher attraction strength leads to faster structure formation.

In this study we have shown how the tuning of basic parameters can influence the structural and dynamical behavior of gel networks. We strengthen the assumption that distinct pathways in gelation, governed by the packing fraction of the network, have major influence on the structural formation as well on a global as on a local level. As we have extended experimental work by Tanaka et al.\cite{tanaka} with our computer simulations, it would be insightful to see how our results compare to further experiments. In addition, the influence of the observed phenomena on mechanical or rheological properties of the gel network might be interesting. Finally, gels usually are evolving very slowly and their formation differs from the behavior observed in other slowly relaxing systems like colloidal glasses thus they represent a separate class of non-equilibrium soft-matter systems \cite{tanaka,schmiedeberg2}. Therefore, connecting our findings on the different types of structural and dynamical behavior to different characterizations of the non-equilibrium will be an important task for future research.
\section*{Data Availability Statement}
The data that support the findings of this study are available from the corresponding author upon reasonable request.

\section*{Conflicts of interest}
There are no conflicts to declare.

\section*{Acknowledgements}
We thank Stefan Egelhaaf for important discussions contributing to this article.

%%%END OF MAIN TEXT%%%

%The \balance command can be used to balance the columns on the final page if desired. It should be placed anywhere within the first column of the last page.

\balance

%If notes are included in your references you can change the title from 'References' to 'Notes and references' using the following command:
%\renewcommand\refname{Notes and references}

%%%REFERENCES%%%
%\bibliography{rsc} %You need to replace "rsc" on this line with the name of your .bib file
%\bibliographystyle{rsc} %the RSC's .bst file

% Create the reference section using BibTeX:
%\bibliography{literature}

\providecommand*{\mcitethebibliography}{\thebibliography}
\csname @ifundefined\endcsname{endmcitethebibliography}
{\let\endmcitethebibliography\endthebibliography}{}

\end{document}